\pdfoutput=1
\documentclass[sigconf]{acmart}

\usepackage[skip=0pt]{caption}
\usepackage[skip=0pt]{subcaption}
\usepackage{multirow}
\usepackage{microtype}
\usepackage{multicol}
\usepackage{tabularx}
\usepackage{graphics}
\usepackage{adjustbox}
\usepackage{nicefrac}
\usepackage{booktabs}

\usepackage{pgfplots}
\usetikzlibrary{pgfplots.groupplots}
\pgfplotsset{compat=1.11}
\usepgfplotslibrary{ternary}
\usepackage{tikz}
\usepackage{tkz-graph}
\usetikzlibrary{positioning,automata}
\usetikzlibrary{calc,spy,shapes}
\usetikzlibrary{arrows,decorations.pathmorphing,fit,positioning,patterns}
\pgfplotsset{compat=newest}

\usepackage{xspace} 

\def\:{\hskip0pt} 

\newcounter{todocnt}

\DeclareFontFamily{U}{mathb}{\hyphenchar\font45}
\DeclareFontShape{U}{mathb}{m}{n}{
<-6> mathb5 <6-7> mathb6 <7-8> mathb7
<8-9> mathb8 <9-10> mathb9
<10-12> mathb10 <12-> mathb12
}{}
\DeclareSymbolFont{mathb}{U}{mathb}{m}{n}

\DeclareMathSymbol{\smalltriangleup} {2}{mathb}{"98}
\DeclareMathSymbol{\smalltriangledown} {2}{mathb}{"99}
\DeclareMathSymbol{\smalltriangleleft} {2}{mathb}{"9A}
\DeclareMathSymbol{\smalltriangleright}{2}{mathb}{"9B}
\DeclareMathSymbol{\blacktriangleup} {2}{mathb}{"9C}
\DeclareMathSymbol{\blacktriangledown} {2}{mathb}{"9D}
\DeclareMathSymbol{\blacktriangleleft} {2}{mathb}{"9E}
\DeclareMathSymbol{\blacktriangleright}{2}{mathb}{"9F}
\copyrightyear{2017}
\acmYear{2017}
\setcopyright{rightsretained}
\acmConference{SIGIR 2017 Workshop on Neural Information Retrieval (Neu-IR'17)}{}{August 7--11, 2017, Shinjuku, Tokyo, Japan} \acmPrice{}

\begin{document}

\title{Share your Model instead of your Data: \\ Privacy Preserving Mimic Learning for Information Retrieval}
\title{Share your Model instead of your Data: \\ Privacy Preserving Mimic Learning for Ranking}
\renewcommand{\shorttitle}{}

\author{Mostafa Dehghani}
\affiliation{%
  \institution{University of Amsterdam}
}
\email{dehghani@uva.nl}

\author{Hosein Azarbonyad}
\affiliation{%
  \institution{University of Amsterdam}
}
\email{h.azarbonyad@uva.nl}

\author{Jaap Kamps}
\affiliation{%
  \institution{University of Amsterdam}
}
\email{kamps@uva.nl}

\author{Maarten de Rijke}
\affiliation{%
  \institution{University of Amsterdam}
}
\email{derijke@uva.nl}

\renewcommand{\shortauthors}{M. Dehghani et al.}

\newcommand{\maingoal}{}
\newcommand{\rqone}{Can we use mimic learning to train a neural ranker?}
\newcommand{\rqtwo}{Are privacy preserving mimic learning methods effective for training a neural ranker?}

\begin{abstract}
Deep neural networks have become a primary tool for solving problems in many fields. They are also used for addressing information retrieval problems and show strong performance in several tasks.  
Training these models requires large, representative datasets and for most IR tasks, such data contains sensitive information from users. 
Privacy and confidentiality concerns prevent many data owners from sharing the data, thus today the research community can only benefit from research on large-scale datasets in a limited manner. 

In this paper, we discuss \emph{privacy preserving mimic learning}, i.e., using predictions from a privacy preserving trained model instead of labels from the original sensitive training data as a supervision signal. 
We present the results of preliminary experiments in which we apply the idea of mimic learning and privacy preserving mimic learning for the task of document re-ranking as one of the core IR tasks.
This research is a step toward laying the ground for enabling researchers from data-rich environments to share knowledge learned from actual users' data, which should facilitate research collaborations. 
\end{abstract}

\keywords{Deep learning; Mimic learning; Responsible information retrieval; Privacy; Model sharing; Data sharing}
%
%



\maketitle

\section{Introduction}
Deep neural networks demonstrate undeniable success in several fields and employing them is taking off for information retrieval problems~\citep{onal2016getting, mitra2017neural}. 
It has been shown that supervised neural network models perform better as the training dataset grows bigger and becomes more diverse~\citep{sun2017revisiting}. 
Information retrieval is an experimental and empirical discipline, thus, having access to large-scale real datasets is essential for designing effective IR systems.
However, in many information retrieval tasks, due to the sensitivity of the data from users and privacy issues, not all researchers have access to large-scale datasets for training their models.

Much research has been done on the general problem of preserving the privacy of sensitive data in IR applications, where the question is how should we design effective IR systems without damaging users' privacy? 
One of the solutions so far is to anonymize the data and try to hide the identity of users~\citep{Carpineto:2013, Zhang:2016}.  As an example, \citet{Zhang:2016} use a differential privacy approach for query log anonymization. However, there is no guarantee that the anonymized data will be as effective as the original data.

Using machine learning-based approaches, sharing the trained model instead of the original data has turned out to be an option for transferring knowledge~\citep{Papernot:2017,Shokri:2015,Abadi:2016}. 
The idea of \emph{mimic learning} is to use a model that is trained based on the signals from the original training data to annotate a large set of unlabeled data and use these labels as training signals for training a new model. 
It has been shown, for many tasks in computer vision and natural language processing, that we can transfer knowledge this way and the newly trained models perform as well as the model trained on the original training data~\citep{Bucilua:2006,Hinton:2015,Romero:2014,Ba:2014}.

However, trained models can expose the private information from the dataset they have been trained on~\citep{Shokri:2015}. Hence, the problem of preserving the privacy of the data is changed into the problem preserving the privacy of the model.
Modeling privacy in machine learning is a challenging problem and there has been much research in this area. Preserving the privacy of deep learning models is even more challenging, as there are more parameters to be safeguarded~\citep{Phan:2016}. 
Some work has studied the vulnerability of deep neural network as a service, where the interaction with the model is only via an input-output black box~\citep{Tramer:2016, Fredrikson:2015, Shokri:2016}.
Others have proposed approaches to protect privacy against an adversary with a full knowledge of the training mechanism and access to the model's parameters. For instance, \citet{Abadi:2016} propose a privacy preserving stochastic gradient descent algorithm offering a trade-off between utility and privacy. More recently, \citet{Papernot:2017} propose a semi-supervised method for transferring the knowledge for deep learning from private training data. They propose a setup for learning privacy-preserving student models by transferring knowledge from an ensemble of teachers trained on disjoint subsets of the data for which privacy guarantees are provided.   

We investigate the possibility of mimic learning for document ranking and study techniques aimed at preserving privacy in mimic learning for this task. Generally, we address two research questions:
\begin{enumerate}
  \setlength{\topsep}{0pt}
  \setlength{\partopsep}{0pt}
  \setlength{\itemsep}{0pt}
  \setlength{\parskip}{0pt}
  \setlength{\parsep}{0pt}
\item[\textbf{RQ1}] \textsl{\rqone}
\item[\textbf{RQ2}] \textsl{\rqtwo}
\end{enumerate}

Below, we first assess the general possibility of exploiting mimic learning for document ranking task regardless of the privacy concerns. 
Then we examine the model by~\citet{Papernot:2017} as a privacy preserving technique for mimic learning.

\section{Training a Neural Ranker with Mimic Learning}
In this section, we address our first research question: ``\rqone''

The motivation for mimic learning comes from a well-known property of neural networks, namely that they are universal approximators, i.e., given enough training data, and a deep enough neural net with large enough hidden layers, they can approximate any function to an arbitrary precision~\citep{Bucilua:2006}. 
The general idea is to train a very deep and wide network on the original training data which leads to a big model that is able to express the structure from the data very well; such a model is called a \emph{teacher} model. 
Then the teacher model is used to annotate a large unlabeled dataset. This annotated set is then used to train a neural network which is called a \emph{student} network. 
For many applications, it has been shown that the student model makes predictions similar to the teacher model with nearly the same or even better performance~\citep{Romero:2014,Hinton:2015}. 
This idea is mostly employed for compressing complex neural models or ensembles of neural models to a small deployable neural model~\citep{Bucilua:2006,Ba:2014}. 

We have performed a set of preliminary experiments to examine the idea of mimic learning for the task of document ranking. 
The question is: Can we use a trained neural ranker on a set of training data to annotate unlabeled data and train a new model (another ranker) on the newly generated training data that works nearly as good as the original model?

\begin{figure}[t]
    \centering
    \includegraphics[height=5cm]{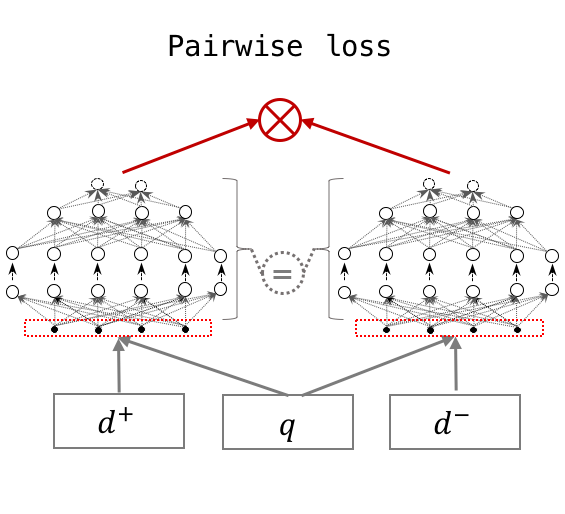}%
    \caption{\label{fig:rankmodel}\emph{Rank Model}: Neural Ranking model proposed by~\citet{Dehghani:2017} }
\end{figure}

In our experiments, as the neural ranker, we have employed \emph{Rank model} proposed by \citet{Dehghani:2017}. The general scheme of this model is illustrated in~\eqref{fig:rankmodel}. 
In this model, the goal is to learn a scoring function $\mathcal{S}(q, d; \theta)$ for a given pair of query $q$ and document $d$ with the set of model parameters $\theta$. 
This model uses a pair-wise ranking scenario during training in which there are two point-wise networks that share parameters and their parameters get updated to minimize a pair-wise loss.
Each training instance has five elements $\tau = (q,d_1, d_2, s_{q,d_1}, s_{q,d_2})$, where $s_{q,d_i}$ indicates the relevance score of $d_i$ with respect to $q$ from the ground-truth.
During inference, the trained model is treated as a point-wise scoring function to score query-document pairs.

In this model, the input query and documents are passed through a representation learning layer, which is a function $i$ that learns the representation of the input data instances, i.e. $(q, d^+, d^-)$, and consists of three components: (1) an embedding function $\varepsilon: \mathcal{V} \rightarrow \mathbb{R}^{m}$ (where $\mathcal{V}$ denotes the vocabulary and $m$ is the number of embedding dimensions), (2) a weighting function $\omega: \mathcal{V} \rightarrow \mathbb{R}$, and (3) a compositionality function $\odot: (\mathbb{R}^{m}, \mathbb{R})^n \rightarrow \mathbb{R}^{m}$. More formally, the function $i$ is defined as:
\begin{equation}
     \begin{aligned}
i(q, d^+, d^-) = [ & \odot_{i=1}^{|q|}(\varepsilon(t_i^q),\omega(t_i^q)) \parallel
& \\ &
\odot_{i=1}^{|d^+|} (\varepsilon(t_i^{d^+}),\omega(t_i^{d^+})) \parallel
& \\ &
\odot_{i=1}^{|d^-|} (\varepsilon(t_i^{d^-}),\omega(t_i^{d^-})) ~],
     \end{aligned}     
\end{equation}
where $t_i^q$ and $t_i^d$ denote the $i$-{th} term in query $q$ and document $d$, respectively.
The weighting function $\omega$ assigns a weight to each term in the vocabulary.
It has been shown that $\omega$ simulates the effect of inverse document frequency (IDF), which is an important feature in information retrieval~\cite{Dehghani:2017}.
The compositionality function $\odot$ projects a set of $n$ embedding-weighting pairs to an $m$-\:dimensional representation, independent of the value of $n$ by taking the element-wise weighted sum over the terms' embedding vectors.
We initialize the embedding function $\varepsilon$ with word2vec embeddings~\cite{Mikolov:2013} pre-trained on Google News, and the weighting function $\omega$ with IDF.

\begin{figure*}[t]
    \centering
    \includegraphics[height=5.5cm]{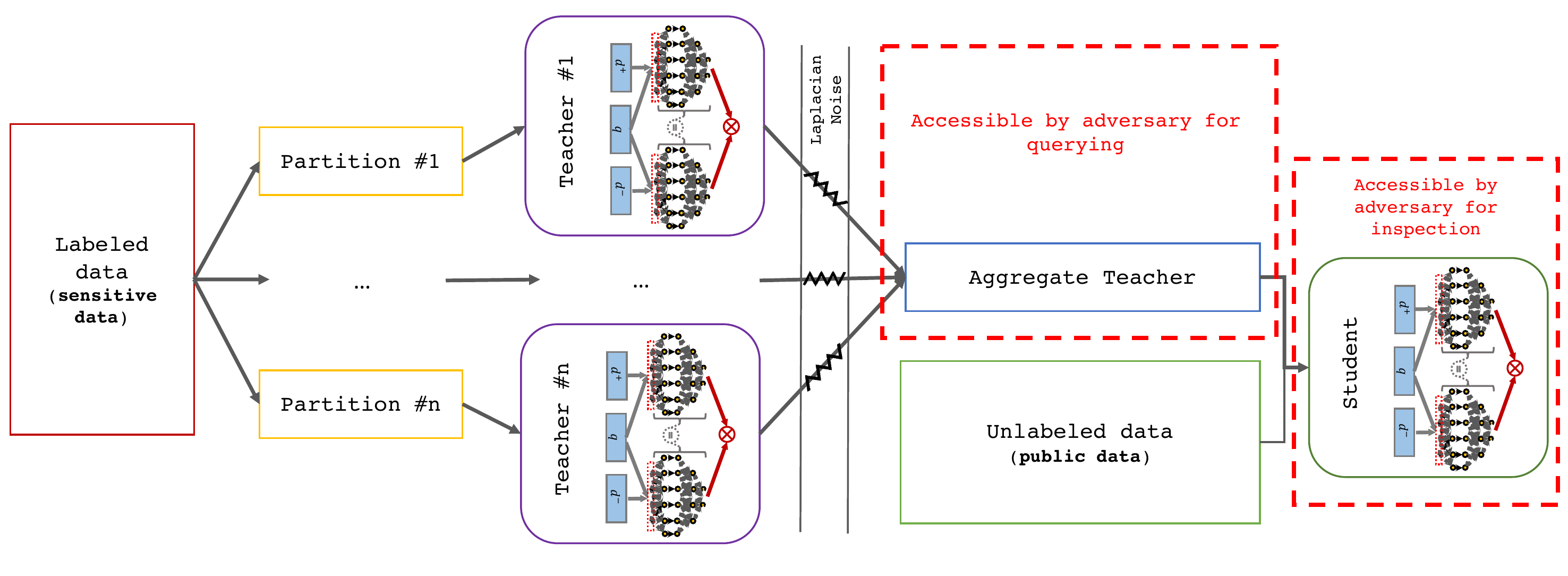}%
    \caption{\label{fig:pp_model} Privacy preserving annotator/model sharing, proposed by~\citet{Papernot:2017}.}
\end{figure*}

The representation learning layer is followed by a simple feed-forward neural network that is composed of $l-1$ hidden layers with ReLU as the activation function, and output layer $z_l$. 
The output layer $z_l$ is a fully-connected layer with a single continuous output with $\tanh$ as the activation function. 
The model is optimized using the hinge loss (max-margin loss function) on batches of training instances and it is defined as follows:
\begin{equation}
\begin{aligned}
\mathcal{L}(b; \theta) = \frac{1}{|b|}
\sum_{i=1}^{|b|}
\max\big\{
& 
0, 1 - \text{sign}
(s_{\{q, d_1\}_i} - s_{\{q, d_2\}_i})
& \\ & 
\left(\mathcal{S}\left(\{q, d_1\}_i; \theta\right) -\mathcal{S}\left(\{q, d_2\}_i; \theta\right)\right)
\big\}
, 
\end{aligned}     
\end{equation}
This model is implemented using TensorFlow~\citep{tang2016:tflearn,tensorflow2015-whitepaper}.
The configuration of teacher and student networks is presented in Table~\ref{tbl:cfg}.

\begin{table}[h]
\centering
\caption{Teacher and student neural networks configurations.}
\vspace{5pt}
\begin{tabularx}{\linewidth}{Xcc} 
\toprule
\bf Parameter & \bf Teacher & \bf Student  \\
\midrule
 Number of hidden layers & 3 & 3  \\
 Size of hidden layers & 512 & 128 \\
 Initial learning rate & 1E-3 & 1E-3 \\
 Dropout & 0.2 & 0.1 \\
 Embedding size & 500 & 300 \\
 Batch size & 512 & 512  \\
\bottomrule
\end{tabularx}
\label{tbl:cfg}
\end{table}

\begin{table}[!h]
\centering
\caption{\label{tbl_res1}Performance of teacher and student models with different training strategies.}
\vspace{5pt}
\begin{adjustbox}{max width=\textwidth}
\begin{tabular}{l l c c c}
\toprule
\bf Training strategy & \bf model & \textbf{MAP} & \textbf{P@20} & \textbf{nDCG@20} 
\\ \midrule
\multirow{2}{*}{{Full supervision}} & {Teacher} 
& 0.1814 & 0.2888 & 0.3419 
\\
& {Student} 
& 0.2256 & 0.3111 & 0.3891 
\\ \midrule
\multirow{2}{*}{{Weak supervision}} & {Teacher} 
& 0.2716 & 0.3664 & 0.4109 
\\ 
& {Student} 
& 0.2701 & 0.3562 & 0.4145 
\\ \bottomrule
\end{tabular}
\end{adjustbox}
\end{table}

As our test collection, we use Robust04 with a set of 250 queries (TREC topics 301--450 and 601--700) with judgments, which has been used in the TREC Robust Track 2004.
We follow the knowledge distillation approach~\cite{Hinton:2015} for training the student network. We have two sets of experiments, in the first one, we train the teacher model with full supervision, i.e., on the set of queries with judgments, using 5-fold cross validation. 
In the second set of experiments, the set of queries with judgments is only used for evaluation and we train the teacher model using the weak supervision setup proposed in~\citep{Dehghani:2017}. We use $3$ million queries from the AOL query log as the unlabeled training query set for the teacher model. 
In all experiments, we use a separate set of $3$ million queries from the AOL query log as unlabeled data that is annotated by the trained teacher model (either using full or weak supervision) for training the student model.

Results obtained from these experiments are summarized in Table~\ref{tbl_res1}. 
The results generally suggest that we can train a neural ranker using mimic learning.
Using weak supervision to train the teacher model, the student model performs as good as the teacher model.
In case of training the teacher with full supervision, as the original training data is small, the performance of the teacher model is rather low which is mostly due to the fact that the big teacher model overfits on the train data and is not able to generalize well. 
However, due to the regularization effect of mimic learning, the student model, which is trained on the predictions by the teacher model significantly outperforms the teacher model \citep{Hinton:2015,Romero:2014}.

\section{Training a Neural Ranker with Privacy Preserving Mimic Learning}
In the previous section, we examined using the idea of mimic learning to train a neural ranker regardless of the privacy risks. In this section, we address our second research question: ``\rqtwo''

It has been shown that there is a risk of privacy problems, both where the adversary is just able to query the model, and where the model parameters are exposed to the adversaries inspection.
For instance, \citet{Fredrikson:2015} show that only by observing the prediction of the machine learning models they can approximately reconstruct part of the training data (model-inversion attack). \citet{Shokri:2016} also demonstrate that it is possible to infer whether a specific training point is included in the model's training data by observing only the predictions of the model (membership inference attack).

We apply the idea of knowledge transfer for deep neural networks from private training data, proposed by \citet{Papernot:2017}. The authors propose a private aggregation of teacher ensembles based on the teacher-student paradigm to preserve the privacy of training data.
First, the sensitive training data is divided into $n$ partitions. Then, on each partition, an independent neural network model is trained as a teacher. 
Once the teachers are trained, an aggregation step is done using majority voting to generate a single global prediction.  
Laplacian noise is injected into the output of the prediction of each teacher before aggregation. The introduction of this noise is what protects privacy because it obfuscates the vulnerable cases, where teachers disagree. 

The aggregated teacher can be considered as a deferentially private API to which we can submit the input and it then returns the privacy preserving label. There are some circumstances where due to efficiency reasons the model is needed to be deployed to the user device~\cite{Abadi:2016}. To be able to generate a shareable model where the privacy of the training data is preserved, \citet{Papernot:2017} train an additional model called the student model. The student model has access to unlabeled public data during training. The unlabeled public data is annotated using the aggregated teacher to transfer knowledge from teachers to student model in a privacy preserving fashion. 
This way, if the adversary tries to recover the training data by inspecting the parameters of the student model, in the worst case, the public training instances with privacy preserving labels from the aggregated teacher are going to be revealed.  The privacy guarantee of this approach is formally proved using differential privacy framework.

We apply the same idea to our task. We use a weak supervision setup, as partitioning the fully supervised training data in our problem leads to very small training sets which are not big enough for training good teachers. 
In our experiments, we split the training data into three partitions, each contains one million queries annotated by the BM25 method. We train three identical teacher models. Then, we use the aggregated noisy predictions from these teachers to train the student network using the knowledge distillation approach. Configurations of teacher and student networks are similar to the previous experiments, as they are presented in Table~\ref{tbl:cfg}.

We evaluate the performance in two situations: In the first one, the privacy parameter, which determines the amount of noise, is set to zero, and in the second one, the noise parameter is set to $0.05$, which guarantees a low privacy risk~\citep{Papernot:2017}.
We report the average performance of the teachers before noise, the performance of noisy and non-noisy aggregated teacher, and the performance of the student networks in two situations.  The results of these experiments are reported in Table~\ref{tbl_res2}.

\begin{table}[h]
\centering
\caption{\label{tbl_res2}Performance of teachers (average) and student models with noisy and non-noisy aggregation.}
\vspace{5pt}
\begin{adjustbox}{max width=\textwidth}
\begin{tabular}{l c c c}
\toprule
 \bf Model & \textbf{MAP} & \textbf{P@20} & \textbf{nDCG@20} 
\\ \midrule
{Teachers (avg)} 
& 0.2566 & 0.3300 & 0.3836
\\ \midrule
{Non-noisy aggregated teacher} 
& 0.2380 & 0.3055 & 0.3702 
\\
{Student \small{(non-noisy aggregation)}} 
& 0.2337 & 0.3192 & 0.3717
\\ \midrule 
{Noisy aggregated teacher} 
& 0.2110 & 0.2868 & 0.3407 
\\
{Student \small{(noisy aggregation)}} 
& 0.2255 & 0.2984 & 0.3559 
\\ \bottomrule
\end{tabular}
\end{adjustbox}
\end{table}

Results in the table suggest that using the noisy aggregation of multiple teachers as the supervision signal, we can train a neural ranker with an acceptable performance.
Compared to the single teacher setup in the previous section, the performance of the student network is not as good as the average performance of teachers. Although the student network performs better than the teacher in the noisy aggregation setup. This is more or less the case for a student together with a non-noisy aggregated teacher.
We believe drops in the performance on the student networks compared to the results in the previous section are not just due to partitioning, noise, and aggregation. This is also the effect of the change in the amount of training data for the teachers in our experiments. So, in the case of having enough training data in each partition for each teacher, their prediction will be more determined and we will have less disagreement in the aggregation phase and consequently, we will get better signals for training the student model.

\section{Conclusion}
With the recent success of deep learning in many fields, IR is also moving from traditional statistical approaches to neural network based approaches.
Supervised neural networks are data hungry and training an effective model requires a huge amount of labeled samples. 
However, for many IR tasks, there are not big enough datasets.
For many tasks such as the ad-hoc retrieval task, companies and commercial search engines have access to large amounts of data. However, sharing these datasets with the research community raises concerns such as violating the privacy of users.
In this paper, we acknowledge this problem and propose an approach to overcome it. Our suggestion is based on the recent success on mimic learning in computer vision and NLP tasks. Our first research question was: \rqone

To answer this question, we used the idea of mimic learning. Instead of sharing the original training data, we propose to train a model on the data and share the model. The trained model can then be used in a knowledge transfer fashion to label a huge amount of unlabeled data and create big datasets. We showed that a student ranker model trained on a dataset labeled based on predictions of a teacher model, can perform almost as well as the teacher model. 
This shows the potential of mimic learning for the ranking task which can overcome the problem of lack of large datasets for ad-hoc IR task and open-up the future research in this direction.

As shown in the literature, even sharing the trained model on sensitive training data instead of the original data cannot guarantee the privacy. Our second research question was: \rqtwo

To guarantee the privacy of users, we proposed to use the idea of privacy preserving mimic learning. We showed that using this approach, not only the privacy of users is guaranteed, but also we can achieve an acceptable performance.
In this paper, we aim to lay the groundwork for the idea of sharing a privacy preserving model instead of sensitive data in IR applications. This suggests researchers from industry share the knowledge learned from actual users' data with the academic community that leads to a better collaboration of all researchers in the field. 

As a future direction of this research, we aim to establish formal statements regarding the level of privacy that this would entail using privacy preserving mimic learning and strengthen this angel in the experimental evaluation. Besides, we can investigate that which kind of neural network structure is more suitable for mimic learning for the ranking task.

\begin{acks}
This research was supported in part by Netherlands Organization for Scientific Research through the \textsl{Exploratory Political Search} project (ExPoSe, NWO CI \# 314.99.108), by the Digging into Data Challenge through the \textsl{Digging Into Linked Parliamentary Data} project (DiLiPaD, NWO Digging into Data \# 600.006.014).

This research was also supported by
Ahold Delhaize,
Amsterdam Data Science,
the Bloomberg Research Grant program,
the Criteo Faculty Research Award program,
the Dutch national program COMMIT,
Elsevier,
the European Community's Seventh Framework Programme (FP7/2007-2013) under
grant agreement nr 312827 (VOX-Pol),
the Microsoft Research Ph.D.\ program,
the Netherlands Institute for Sound and Vision,
the Netherlands Organisation for Scientific Research (NWO)
under pro\-ject nrs
612.001.116, 
HOR-11-10, 
CI-14-25, 
652.\-002.\-001, 
612.\-001.\-551, 
652.\-001.\-003, 
and
Yandex.
All content represents the opinion of the authors, which is not necessarily shared or endorsed by their respective employers and/or sponsors.

\end{acks}

\bibliographystyle{ACM-Reference-Format}
\bibliography{ref} 
\end{document}